%% file: BHclass.tex
\documentclass[floats,floatfix,showpacs,amssymb,prd,twocolumn,superscriptaddress,nofootinbib,nolongbibliography,reprint]{revtex4-2}

\usepackage{amssymb,amsmath,verbatim,mathtools,needspace,enumitem,etoolbox,graphicx,physics,microtype,afterpage,bigints,gensymb,tabularx,mathtools,siunitx, scalerel}

\usepackage[dvipsnames, usenames]{xcolor}
\definecolor{linkcolor}{rgb}{0.0,0.3,0.5}
\definecolor{dodgerblue}{HTML}{1E90FF}
\usepackage{hyperref}
\hypersetup{unicode, colorlinks=true, linkcolor=linkcolor, citecolor=linkcolor, filecolor=linkcolor,urlcolor=linkcolor, pdfusetitle}
\usepackage[all]{hypcap}
\usepackage[T1]{fontenc}
\usepackage[utf8]{inputenc}
\usepackage{orcidlink}

\usepackage [english]{babel}
\usepackage [autostyle, english = american]{csquotes}
\MakeOuterQuote{"}

\usepackage{soul}
\usepackage{bbm}
\usepackage{enumitem}
\usepackage{multirow}

\input{journal_macros.tex}

\renewcommand{\emph}[1]{\textit{#1}}

\newcommand{\umani}{\affiliation{Department of Physics and Astronomy \& Winnipeg Institute for Theoretical Physics, University of Manitoba, Winnipeg, Manitoba, R3T 2N2, Canada}}

\newcommand{\Msol}{\rm \,M_{\odot}}

\begin{document}

\title{Machine learning classification of black holes in the mass--spin diagram}

\author{Nathan Steinle$\,$\orcidlink{0000-0003-0658-402X}}
\email{nathan.steinle@umanitoba.ca}
\umani

\author{Samar Safi-Harb$\,$\orcidlink{0000-0001-6189-7665}}
\umani

\begin{abstract}
We present the mass--spin diagram for classifying black holes and studying their formation pathways providing an analogue to the Hertzsprung–Russell diagram. 
This allows for black hole evolutionary tracks as a function of redshift, combining formation, accretion, and merger histories for the variety of black hole populations. 
A realistic black hole continuum constructed from initial mass and spin functions and approximate redshift evolution reveals possible black hole main sequences, such as sustained coherent accretion through cosmic time or hierarchical merger trees. 
In the stellar-mass regime, we use a binary population synthesis software to compare three spin prescriptions for tidal evolution of Wolf-Rayet progenitors, showing how the mass--spin diagram exposes interesting modeling differences. 
We then classify black hole populations by applying supervised and unsupervised machine learning clustering methods to mass--spin datasets. 
While bare unsupervised clustering can nearly recover canonical population boundaries (stellar-mass, intermediate-mass, and supermassive), a more  sophisticated approach utilizing deep learning via variational autoencoders for latent space representation learning aids in clustering of  realistic datasets with subclasses that highly overlap in mass--spin space. 
We find that a supervised random forest can accurately recover the correct clusters from the learned latent space representation depending on the complexity of the underlying dataset, semi-supervised methods show potential for further development, and the performance of unsupervised classifiers is a great challenge.   
Our findings motivate future machine learning applications and demonstrate that the mass--spin diagram can be used to connect gravitational-wave and electromagnetic observations with theoretical models. 
\end{abstract}

\maketitle

\section{Introduction} 
\label{sec:Intro} 

The Hertzsprung–Russell (HR) diagram \cite{1914PA.....22..331R,1911POPot..63.....H} is a foundational tool in astronomy \cite{2014A&A...564A..52L,2017imas.book.....C,2018A&A...616A..10G}. Serving as a quantitative bridge for understanding stellar structure, evolution, and classification, it provides a canvas for correlations between a star's luminosity and surface temperature. The HR diagram revealed underlying patterns of observational data and motivated new models of stars. 

Modern stellar astrophysics builds upon this diagram with detailed models of stellar interiors, implemented in various codes ranging from detailed simulations to advanced numerical tools and comprehensive rapid binary population synthesis models that rely on fits to the more detailed models. All such models can be used to generate theoretical HR diagrams by constructing populations based on evolutionary modeling. 

\begin{figure}
\centering
\includegraphics[width=0.48\textwidth]{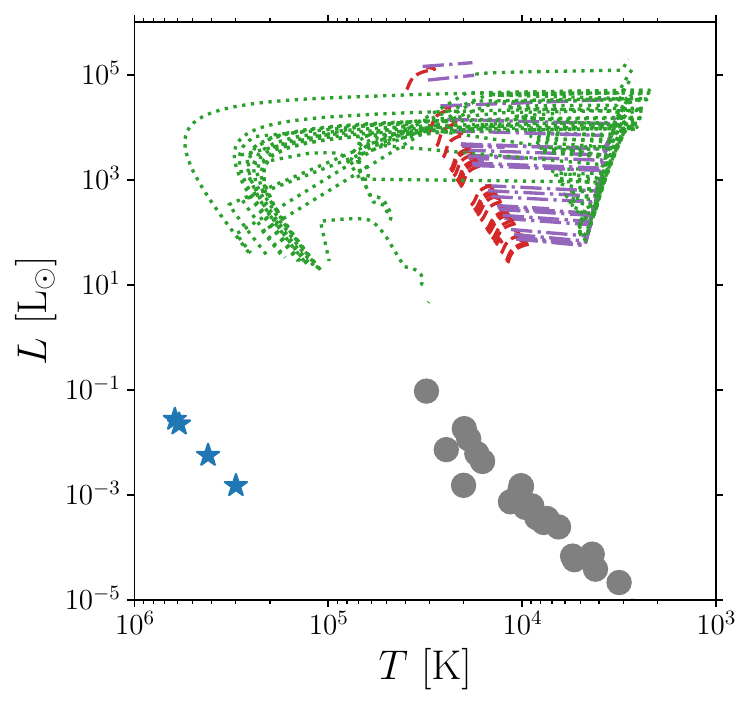}
\caption{
A Hertzsprung-Russell diagram (luminosity vs effective surface temperature) showing the evolutionary tracks of a handful of stars modeled with the COMPAS binary population synthesis software \cite{COMPAS2022} through the stellar main sequence and Hertzsprung gap (red dashed line), giant branches (purple dash-dot line), and core helium burning (green dotted line) phases. Some of these stars form into white dwarfs (gray dots) or neutron stars (blue stars). 
} \label{F:HRdiagram}
\end{figure}

Figure~\ref{F:HRdiagram} displays an HR diagram from a synthetic stellar population generated with the \textsc{COMPAS}\footnote{\href{https://compas.science/}{COMPAS} is a publicly available population synthesis code designed for evolution prescriptions of isolated binary-star systems.} isolated-binary evolution code with default parameter settings \cite{COMPAS2022}. 
The stellar classifications, as the HR diagram reveals, are intimately tied to the phases of stellar evolution through the stars' lifespans. 
We filter this prototypical theoretical population to obtain a small randomized subset whose evolutionary stages are shown in Fig.~\ref{F:HRdiagram}. 
In chronological order, these intermediate-to-high mass stars begin core hydrogen fusion, i.e. the main sequence (red dashed lines) and move through a very brief Hertzsprung gap characterized by shell helium burning, which is followed by drastic radial expansion as core helium fusion ignites and constitutes the dominant component of the stellar luminosity in the giant branch (dot-dash purple lines). Then these stars lose their envelopes, either through binary mass transfer processes or stellar wind mass loss, revealing the core-helium burning phase of stellar evolution (dotted green lines) which leads imminently to gravitational collapse, possibly associated with a supernova explosion and compact object formation. A few surviving white dwarfs (gray dots) and neutron stars (blue stars) occupy their respective lower luminosity regions with the neutron stars effective temperature being much higher due to differences in their supernovae mechanisms. This highlights the utility of the HR diagram for organizing the variety of stellar types and diagnosing stellar evolution models. 
Extensions of the HR diagram are still being found for varied purposes \cite{1995ARA&A..33...75G,1999MNRAS.304..705H,2013A&A...554A.108M,2014A&A...564A..52L,2024MNRAS.528.4272H}, underscoring its impact in modern astronomy. 
However, black holes could only be drawn on the HR diagram if they are embedded in an accretion disk or due to non-classical possibilities such as Hawking radiation. 

Motivated by this paradigm, we ask: can an analogous diagram be constructed for astrophysical black holes? 
We propose that, similar to the HR diagram, one can visually explore how black holes can be organized and classified. 
While black holes are historically challenging to observe, their known macroscopic properties of mass and spin from General Relativity offer natural axes for classification. 
With electromagnetic (EM) techniques such as X-ray spectroscopy and radio interferometry providing constraints on stellar-mass ($\approx 5 - 10^2 \Msol$), intermediate-mass ($\approx 10^2 - 10^5 \Msol$), and supermassive ($\gtrsim 10^6 \Msol$) black holes \cite{2017RvMP...89b5001B,2015PhR...548....1M,2013CQGra..30x4004R,2023arXiv231016896D,2007ApJ...656...84G,2019ApJ...882..181B} and with gravitational-wave (GW) detections now delivering catalogs of stellar-mass black hole mergers \cite{2019PhRvX...9c1040A,2021PhRvX..11b1053A,2023PhRvX..13d1039A}, we can consider how patterns in black hole properties may emerge. 

\begin{figure*}
\centering
\includegraphics[width=\textwidth]{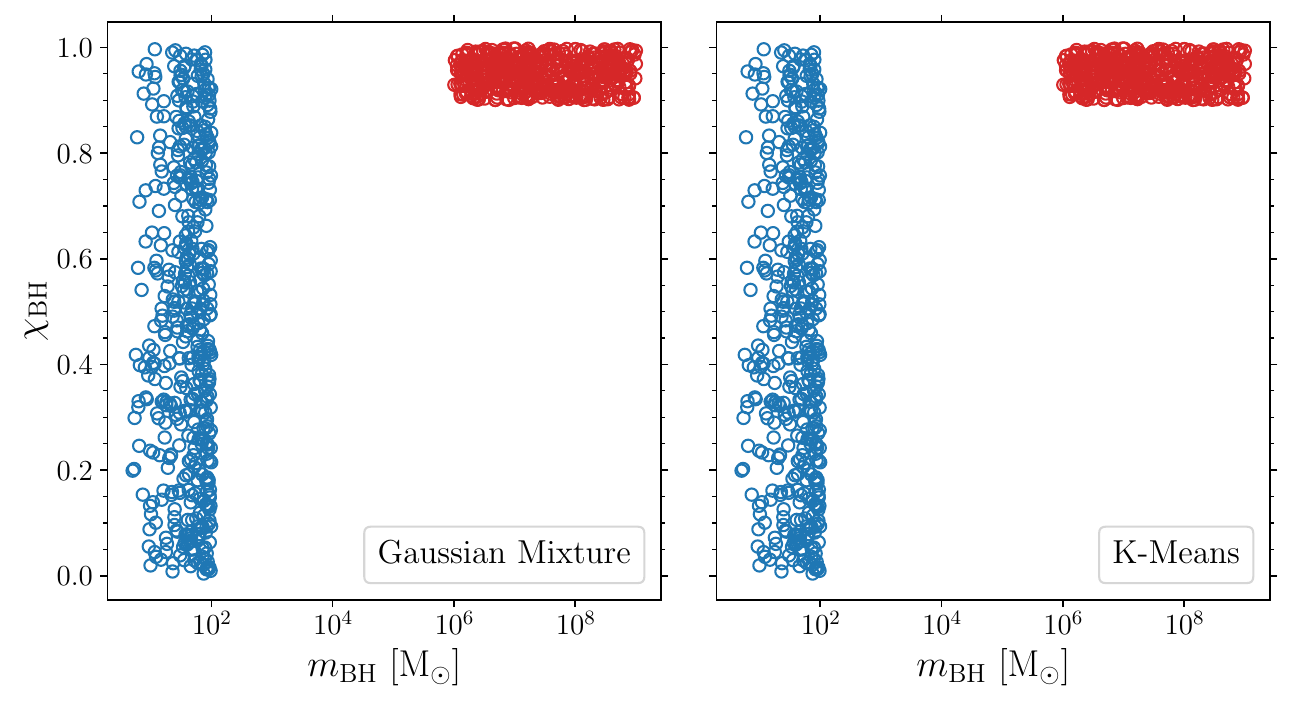}
\caption{
A simple example of a black hole mass gap, i.e. no intermediate-mass black holes, and distinct and separated mass--spin correlations for stellar-mass and supermassive black holes. Given to bare unsupervised clustering algorithms \cite{2011JMLR...12.2825P}, k-means (right panel) and a Gaussian mixture model (left panel) cluster with 100\% accuracy. 
} \label{F:IMBHgap}
\end{figure*}

Recent years have seen substantial effort in quantifying and interpreting the mass--spin relations of BHs across the mass spectrum. For stellar‑mass black holes in X‑ray binaries, two primary techniques—continuum‑fitting and iron‑line reflection spectroscopy—have yielded a range of spin magnitudes, underscoring the complexity of natal spin and accretion histories \cite{2006AN....327..997M,2011MNRAS.414.1183K}. At the supermassive end, X‑ray reflection analyses show that the majority with masses below $\approx\num{3e7} \Msol$ are rapidly spinning ($\chi \gtrsim$ 0.9), while those with higher masses exhibit a tentative trend toward lower spins \cite{2013CQGra..30x4004R}. The elusive intermediate‑mass black holes remain less certain, but numerous strong candidates are found in globular clusters, hyper‑luminous X‑ray sources, and dwarf-galaxy nuclei, albeit with large uncertainties \cite{2017IJMPD..2630021M,2020ARA&A..58..257G}, and a few have been confidently detected via GWs, such as GW190521 \cite{2020PhRvL.125j1102A} and GW231123 \cite{2025arXiv250708219T}, calling into question the mass boundaries that typify this mass regime. 
These and other classes of black holes are summarized in Table~\ref{Table:summary_BHs} along with their observed mass and spin ranges which we use to construct the datasets used in Fig.'s \ref{F:redshift}, \ref{F:IMBHgap}, \ref{F:continuum}, \ref{F:latent} and in Table~\ref{Table:summary_stats_case1} of Sections \ref{sec:BHMassSpinDiagram} and \ref{sec:Classifying}. 

\input{table.tex}

Emergent gravitational‑wave population studies suggest a mass‑dependent spin distribution in merging binaries \cite{2024A&A...692A..80P}, and prior studies have explored black hole mass, spin, and redshift relationships in GW data \cite{2018ApJ...868..140T,2022ApJ...928..155T,2024A&A...684A.204R}. An observed anti-correlation in the $\chi_{\rm eff}$--$q$ plane \cite{2020ApJ...894..129S,2021ApJ...922L...5C,2021PhRvD.104h3010R}, where $\chi_{\rm eff}$ is the aligned effective spin parameter \cite{2001PhRvD..64l4013D,2008PhRvD..78d4021R} and $q$ the mass ratio, was subsequently confirmed in more recent analyses of gravitational-wave catalogs and helps reveal subpopulations in the observed dataset associated with different formation scenarios \cite{2023ApJ...958...13A,2023PhRvX..13a1048A,2025ApJ...987...65L}. 
Structures can be inferred from the spin distributions using data-driven models that capture unanticipated features without inserting systematic bias \cite{2023PhRvD.108j3009G}. 
Investigations into the influence of stellar rotation on black hole mass--spin outcomes has long been studied and is model dependent \cite{2005A&A...429..581M,2024MNRAS.534.1868G}, which often involves implicitly projecting onto the mass--spin plane. 
These works significantly advance understanding of stellar-mass black hole populations, and motivate development of a tool for interpreting black hole mass and spin measurements. 

To begin sketching this framework, we present a cartoon schematic of the mass--spin plane 
containing a 
feature labeled ``Cosmic Accretion'' that represents 
a clustering of black holes, 
analogous to the stellar main sequence, defined by extended coherent accretion episodes over cosmic time. While there is only one main sequence for stars, i.e., the dominant band of stars on the HR diagram where they spend most of their lives, we propose that black holes may have many depending on if and how they grow in mass and spin. 
We quantify this by sampling from physically motivated mass and spin distributions across the stellar-mass, intermediate-mass, and supermassive black hole regimes, and modeling their redshift evolution under Eddington-limited accretion to trace evolutionary tracks/trajectories across the diagram. 
Next, we construct a mass--spin diagram focused on the isolated-binary formation channel \cite{2014LRR....17....3P} of stellar-mass black holes using three state-of-the-art prescriptions for obtaining high black hole spin magnitude from tidal synchronization of its Wolf-Rayet progenitor's spin, demonstrating the model-specific regions that a population of black holes can occupy in a canonical formation scenario. 

We then apply (unsupervised and supervised) machine learning algorithms to many example populations. 
We show that a simple unsupervised clustering approach (with spectral clustering algorithm) can recover clusters of the main black hole mass regimes — stellar-mass, intermediate-mass, and supermassive black holes — but more challenging datasets with overlapping subpopulations require sophisticated methods. 
For three such datasets, we employ deep learning (via  variational autoencoders in \textsc{PyTorch} \cite{paszke2017automatic}) for latent space representation and use clustering algorithms (via \textsc{Sci-kit Learn} \cite{2011JMLR...12.2825P}) to distinguish models of black hole spin evolution and formation. 
The first dataset contains the most distinct subclasses, allowing for accurate clustering by the supervised random forest, and the remaining two datasets are more challenging as their subclasses (i.e., isolated binaries vs dynamical-cluster binaries) exhibit high overlap in the mass--spin space. 
On the other hand, unsupervised clustering is unable to learn the latent space of all three datasets, and we demonstrate that semi-supervised clustering methods show great potential for handling such datasets.
These results lay the groundwork for numerous future studies in black hole classification. 

This paper is organized as follows. We present the theoretical motivation for the black hole mass-spin diagram 
and illustrative examples to show its utility in Sec.~\ref{sec:BHMassSpinDiagram}. In Sec.~\ref{sec:Classifying} we apply machine learning classification pipelines to mass--spin datasets. In Sec.~\ref{sec:Sum} we summarize our analysis and findings, and in Sec.~\ref{sec:Disc} we discuss the implications of our results with emphasis on future investigations. 

\begin{figure*}
\centering
\includegraphics[width=\textwidth]{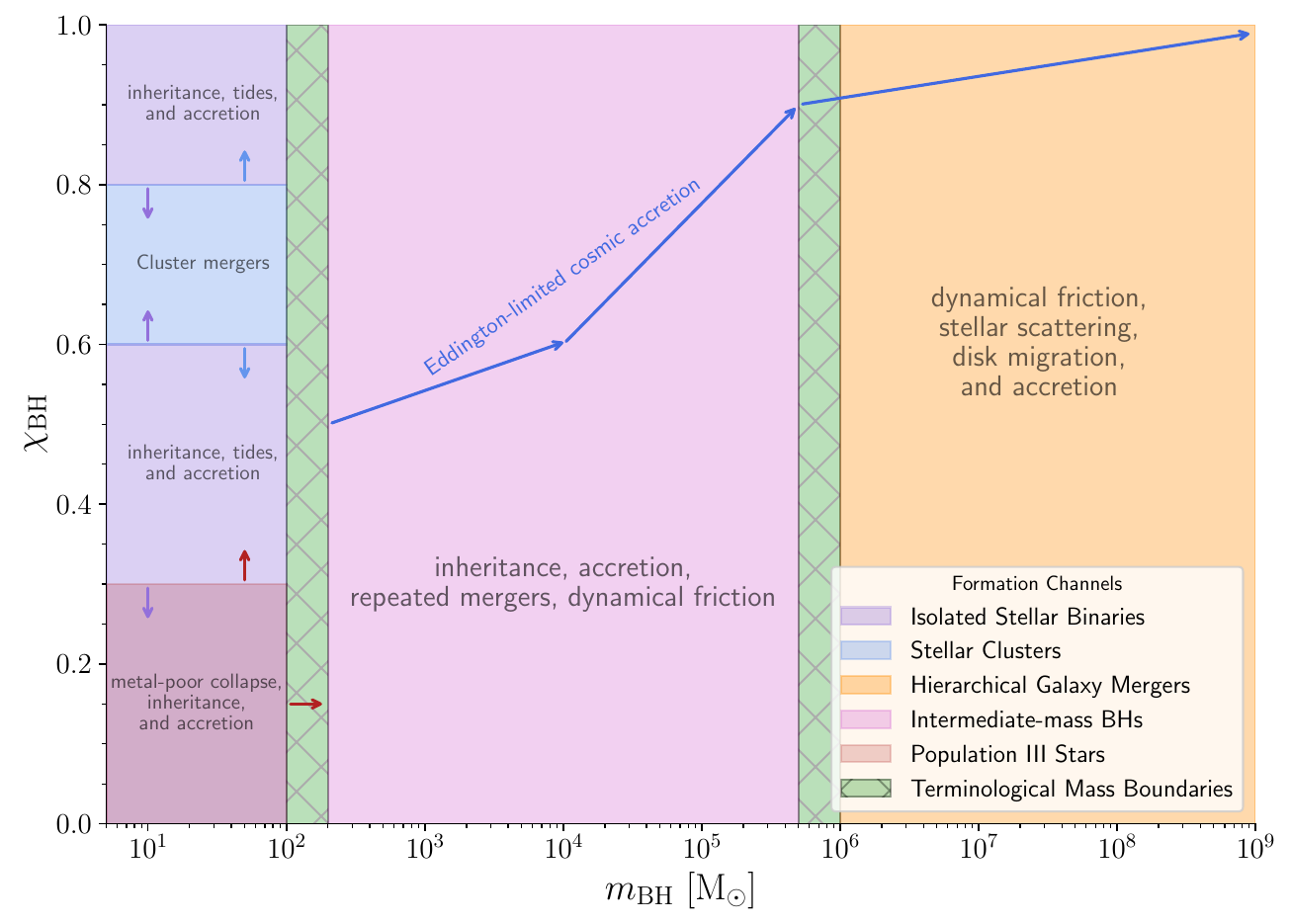}
\caption{
A sketch of the mass--spin diagram for astrophysical black holes with dimensionless spin angular momentum (i.e. $\chi = c|\mathbf{S}|/Gm^2$) on the vertical axis. Regions are colored, as indicated in the legend, according to a selection of formation channels of black holes in single, binary, or higher systems across the mass spectrum: binary stars in isolation, dense stellar clusters, canonical population III stars, and hierarchical galaxy mergers. In reality, these formation channels will 
overlap and their prevalence will depend on redshift. 
The two green vertical patches with red hatch symbolize the commonly used boundaries between the stellar-mass, intermediate-mass, and supermassive regimes. 
Typical astrophysical processes that drive the mass and spin growth across the diagram are written in each regime, with small arrows indicating uncertainty in the region borders for mass and spin predictions of formation channels. The solid blue line extending from the stellar-mass regime to the upper right corner of the diagram represents  
a possible 
main-sequence analogue for black holes from sustained accretion episodes over cosmic time.   
} \label{F:cartoon}
\end{figure*}

\section{Black hole mass-spin diagrams}
\label{sec:BHMassSpinDiagram}

Black holes cannot meaningfully be placed on an HR diagram as they lack intrinsic luminosity but a \textit{mass--spin diagram} provides an analogous classification tool. In this space, every black hole occupies a point whose coordinates can shift over cosmic time due to formation, accretion, merger events, or even Hawking evaporation, thereby tracing out trajectories (or tracks) that reflect astrophysical and cosmological history. 
The mappings between redshift and environment-dependent accretion states, and between pre- and post-merger binary component masses and spins, imply the existence of black hole main sequence analogues for dominant growth pathways of mass and spin. Understanding such mappings allows inference of black hole evolutionary typology across cosmic time. 

From a theoretical standpoint, the \textit{Kerr hypothesis} states that, in classical general relativity, an astrophysical black hole is described by its mass and spin\footnote{Throughout this work, we use the dimensionless spin parameter of the black hole which is the specific angular momentum of the black hole per unit mass, i.e., $\chi \equiv cS/Gm^2$ where $S$ is the spin angular momentum and $c$ and $G$ are the speed of light and gravitational constant, respectively.} \cite{2000CQGra..17.3353M,2005hep.ph...11217H,2011MPLA...26.2453B,2015CQGra..32l4006T}. Of course, this depends on how one defines a black hole \cite{2019NatAs...3...27C}, which can be accomplished from the event horizon without mention of the curvature singularity \cite{1973grav.book.....M}. In practice, inference of black hole properties relies on modeling of environments, such as accretion disks, or gravitational radiation. 
It is often sufficient (and useful) to simplify to a non-rotating (Schwarzschild) black hole, depending on the data or theoretical problem being analyzed. As such, the development of black hole astrophysics has paralleled the rise of sophisticated approximations to black hole spacetimes \cite{2005NJPh....7..204F}, such as perturbation theory and the post-Newtonian formulations of General Relativistic effects which have provided outstanding utility relative to the uncertainties of the strong gravity regime  \cite{2011PNAS..108.5938W}. 

Observationally, mass is the first and easiest parameter to constrain. 
Observations of gravitational waves from coalescences of stellar-mass binaries with the LIGO/Virgo/KAGRA network (LVK) \cite{2019PhRvX...9c1040A,2021PhRvX..11b1053A,2023PhRvX..13d1039A} are most sensitive to a combination of the binary masses known as the chirp mass, and can offer weak constraints on spin magnitudes and directions typically expressed through the effective aligned-spin parameter $\chi_{\rm eff }$ \cite{2008PhRvD..78d4021R,2001PhRvD..64l4013D} and effective precession parameter $\chi_{\rm p}$ \cite{2015PhRvD..91b4043S,2021PhRvD.103f4067G}, but the spin magnitudes can be better measured for certain systems \cite{2014PhRvL.112y1101V,2015arXiv151204955P,2021PhRvL.126q1103B}. 
The binary black hole spin-orbit misalignments, also known as spin tilts or orientations, are expected to carry imprints of their progenitor evolution and hold great implications for general relativistic spin precession, but they are not yet well constrained from GW data \cite{2017MNRAS.471.2801S,2022A&A...668L...2V,2025arXiv250514875V}. 

Modern EM techniques have placed increasingly precise limits on spin magnitudes, both in X-ray binaries and quasars/AGN \cite{2015PhR...548....1M,2013CQGra..30x4004R,2025arXiv250115380M}. Measurements of X-ray binaries can in principle constrain the spin orientation of the accreting black hole but the orbital inclination introduces burdensome systematics. Furthermore, it is an active area of research to develop new methods for EM observations to confidently identify dual- or multi-AGNs \cite{2024arXiv241112799P} and to distinguish a single supermassive black hole from a supermassive black-hole binary within a galactic nucleus \cite{2023arXiv231016896D}, indicating a bright future for EM measurements of black hole spin orientations relative to a companion or other environmental aspects. 

Black holes are observed confidently in two main mass regimes: 
stellar-mass black holes through electromagnetic methods, such as accreting binaries \cite{2015PhR...548....1M,2020mbhe.confE..28B} or astrometry, and through the terrestrial gravitational-wave detector network LVK; and supermassive black holes in galactic nuclei and quasars, observed via multi-wavelength spectroscopy, gas and stellar dynamics, and reverberation mapping \cite{2014SSRv..183..253P,2014SSRv..183..277R,2013CQGra..30x4004R,2018Ap&SS.363...82K,2021iSci...24j2557C,2020mbhe.confE..28B}, and hopefully via gravitational waves with a future space-based detector such as LISA via binary or minor mergers \cite{2017arXiv170200786A}. 
The intermediate-mass regime is populated with many strong candidates \cite{2017IJMPD..2630021M,2020ARA&A..58..257G} but has not yet been confidently confirmed beyond the few hundred solar mass extraordinary events from LVK. While these LVK events confirm the existence of black holes with masses greater than 100 $\Msol$ \cite{2025ApJ...985L..37R,2025arXiv250708219T}, it remains to be seen how the rest of the gap, i.e. from $\sim$1000 to $\sim10^5$, is filled. 

The current nomenclature---\textit{stellar-mass}, \textit{intermediate-mass}, \textit{supermassive}, even \textit{very massive} or \textit{stupendously massive} \cite{2021MNRAS.501.2029C}, as well as \textit{primordial black holes} \cite{2018CQGra..35f3001S}---reflects a combination of formation channels and phenomenological mass ranges. Hawking radiation provides theoretical grounding for the expectation that \textit{micro-black holes} ($M \lesssim 10^{12}\,\mathrm{kg}$) cannot survive to the present epoch (they would have evaporated long ago as their Hawking timescale is $\propto M^3$), implying such objects are unlikely to be astrophysical \citep{1975CMaPh..43..199H,1976PhRvD..13..198P}. 

By combining theoretical predictions, astrophysical modeling of origin scenarios, and observational constraints across mass ranges, the \textit{mass--spin diagram} emerges as a unified framework. Analogous to the stellar HR diagram, it enables us to classify black holes and compare populations by redshift, formation channels, and environs. 

Across the black-hole mass spectrum, several formation channels populate regions or trajectories in the mass--spin plane, which can be distinct or highly overlapping depending on the underlying assumptions. 
Stellar core collapse can produce stellar-mass black holes with low to moderate natal spins depending on the nature of angular momentum transport and possible tidal evolution during progenitor evolution \cite{2019ApJ...881L...1F,2023ApJ...952...53M}. 
Binary mergers in dense environments lead to spin-up and mass growth via well-characterized remnant formulae. 
Accretion-driven evolution from both disk feeding and tidal disruptions can also significantly affect black hole spins depending on the Eddington fraction of the accretor and the tidal radius of the disruptor, respectively. 
Supermassive black holes grow through galaxy mergers and prolonged accretion episodes, resulting in mass--spin evolution that is skewed toward high spin in low-mass supermassive black holes but possibly limiting at high mass \cite{2005ApJ...620...69V,2008ApJ...684..822B,2014MNRAS.440.1590D}. 
Also primordial black holes which, if they exist, could span a wide range of masses and spins and serve as fossils for cosmology.

\begin{figure*}
\centering
\includegraphics[width=\textwidth]{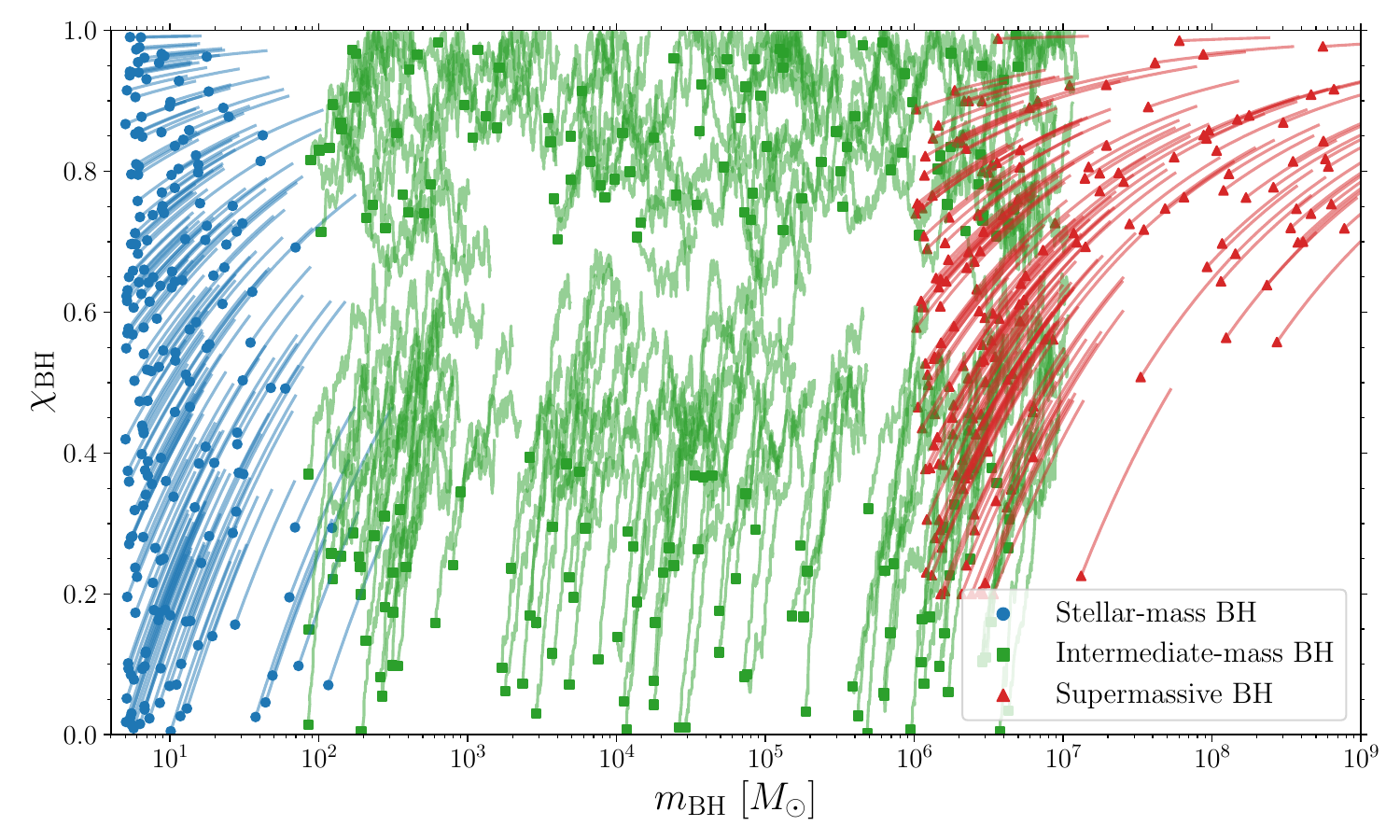}
\caption{
Theoretical evolutionary tracks of accreting black holes across the mass--spin diagram. Black holes are drawn from initial mass functions and simple natal spin prescriptions, and are evolved from redshift $z = 10$ to $z = 0$ along the trajectories. 
Stellar (blue dots and lines) and supermassive (red triangles and lines) black holes undergo an approximate thin disk evolution, while intermediate-mass black holes (green squares and lines) experience chaotic accretion modeled with a Gaussian drift. The clusters that result from the flow of points toward the upper-right corner of the panel is a quantitative example of a \emph{Cosmic Accretion} main sequence (i.e., the blue solid line of Fig.~\ref{F:cartoon}). 
} \label{F:redshift}
\end{figure*}

In Figure~\ref{F:cartoon} we present a schematic mass--spin diagram summarizing some main theoretical possibilities and expected trajectories. Each shaded region corresponds to a specific black hole formation channel, such as isolated stellar binaries, stellar clusters, hierarchical galaxy mergers, or Population\,III stars, and spans a range of spin magnitudes and black hole masses predicted by each scenario. Horizontal and vertical transitions are overlaid to indicate astrophysical processes relevant for growing the black hole masses and/or spins. The processes for intermediate-mass black holes, which may be populated by multiple overlapping formation pathways, are expected to be a mixture of processes identified from the stellar (i.e. stellar collapse, spin inheritance, tidal spin-up of the progenitor, and accretion) and supermassive (dynamical friction, binary loss-cone stellar scattering, disk-driven migration and accretion) regimes \cite{2017IJMPD..2630021M, 2020ARA&A..58..257G}. 
The commonly used boundaries between the three mass regimes are shown for reference in green regions with red hatching. The small arrows colored by formation channel indicate uncertainties on the theoretical boundaries of the formation channel. Therefore, this diagram serves as a visual analog to the HR diagram for stars: a two-dimensional landscape in which different populations occupy distinct, physically motivated regions, and evolutionary tracks and snapshots can be mapped and classified. 
The solid blue line extending across the diagram signifies a conservative upper limit on the spins of black holes undergoing Eddington-limited accretion. 
The flow of tracks toward the upper-right corner 
motivate a black hole main sequence analogue we refer to as \emph{Cosmic Accretion}, i.e., a 
cosmological period of coherent accretion growth. 
We note that super-Eddington accretion sources will produce tracks to higher spins for a given mass. 

The structure of this diagram encodes several key physical insights. Black holes formed from isolated stellar binaries are expected to cluster at low to intermediate masses, with spins shaped by the angular momentum of their progenitor stars and subsequent tidal interactions or accretion \cite{2014LRR....17....3P}. Stellar clusters, which foster dynamical encounters and hierarchical mergers, can produce second-generation black holes with higher spins, populating the intermediate-$\chi$ region \cite{Benacquista2013,2022MNRAS.511.5797M}. 
In the supermassive regime, high spins emerge from coherent gas accretion, merger histories in massive halos, or possibly inheritance during formation. Meanwhile, the metal-poor Population\,III stars offer an alternative low-spin, high-mass origin for stellar black holes, especially in the early universe. The intermediate-mass region likely represents multiple possibilities, including cluster-driven mergers, off-nuclear minor mergers, runaway collapse, and chaotic accretion. Our diagram highlights that, just as stars follow coherent evolutionary paths across the HR diagram, black holes trace mass--spin trajectories tied to their cosmic origin and evolution. This motivates the use of the mass--spin plane as a diagnostic framework for both classification and inference of black hole populations.

We quantitatively illustrate this in Figure~\ref{F:redshift} which shows the redshift-dependent trajectories of black holes across the mass--spin plane under different physical growth scenarios, representing stellar-mass black holes (blue circles), intermediate-mass black holes (green squares), and supermassive black holes (red triangles). Each line, i.e. trajectory, traces the evolution of an individual black hole from redshift $z = 10$ to $z = 0$, with markers indicating the initial position in mass--spin space. The stellar-mass black holes are born with a power-law mass distribution ($m \sim m^{-2.3}$ \cite{1955ApJ...121..161S}) and uniform spin magnitudes \cite{2017PhRvD..95l4046G,2018PhRvD..98h4036G}, and evolve under an approximation to the classic model of \cite{1974ApJ...191..507T} where we assume $z_i \rightarrow z_j$ with $i<j$ defines a redshift timestep. 
In this approximate model, a black hole of mass $m$ accretes an amount of mass $\Delta m(z) = m(z_j) - m(z_i) = m_{\rm n}(1 + 1.5(1 - \bar{z}))$ where $\bar{z} = z / 10$ and $m_{\rm n}$ is the natal mass, and undergoes corresponding spin increase via 
$\Delta \chi (z) = \chi_i + (\chi_{\rm eq} - \chi_i)\eta \Delta m(z)/m$ where $\chi_{\rm eq} = 0.998$ and $\eta = 0.3$ is an ad-hoc parameter that controls the efficiency of spin-up. 
Intermediate-mass black holes are drawn from a log-flat mass distribution and bimodal initial spin distribution \cite{2018ApJ...856...92F}, and undergo modest stochastic mass growth $\Delta m(z) = m_{\rm n}(1 + 1.5(1 - \bar{z}))(1 + \mathcal{N}(0, 0.003))$ accompanied with conditional spin drift (i.e., aggressive upward spin drift if $\chi < 0.6$ at a given $z$), loosely modeling the combined effect of hierarchical mergers and chaotic accretion \cite{2003ApJ...585L.101H,2017IJMPD..2630021M}. 
We assume supermassive black holes originate from a 60\%/40\% mixture of light and heavy seed populations (light seeds from a Schechter-like distribution with natal spins $\sim \mathcal{N}(0.5, 0.2)$ and heavy seeds from a log-uniform distribution with natal spins from $\sim \mathcal{N}(0.8, 0.1)$) \cite{2006MNRAS.371.1813L,2006MNRAS.370..289B,2008ApJ...684..822B,2009NewAR..53...57S,2012MNRAS.423.2533B,2020ARA&A..58...27I,2021NatRP...3..732V}, and grow through coherent thin-disk accretion via $\Delta m(z) = m_{\rm n}(1 + 3.0e^{-\bar{z}/1.5})$ and with efficient spin-up modeled similar to the stellar-mass case. The resulting diagram shows the imprints of these evolutionary pathways in the mass--spin space, which cluster in regions over time and reveal mass--spin gaps (for example in the top-left, middle, and bottom right portions of the diagram) analogous to current mass gaps such as the pair instability and neutron star-black hole gaps. 

The evolutionary tracks reveal key insights into how different astrophysical growth mechanisms shape the mass--spin distribution of black holes over cosmic time. Stellar-mass black holes, assumed to begin with low to moderate spins, can achieve high spin magnitudes ($\chi \gtrsim 0.9$) through sustained super-Eddington accretion \cite{2022ApJ...933...86Z}. However, we do not include the presence of dynamically formed black holes which would cluster at moderate spin. 
Intermediate-mass black hole evolution reflects inefficient or chaotic angular momentum transfer during hierarchical assembly in globular clusters or dwarf galaxies \cite{2004ApJ...602..312G}. Supermassive black holes undergo substantial mass growth and efficient spin-up, with final spins approaching the Thorne limit ($\chi \approx 0.998$), as expected from prolonged disk accretion episodes in high-redshift quasars \cite{2019MNRAS.490.4133B}. Importantly, the structure of this mass--spin plane reveals continuous, physically-driven redshift evolution offering a rich framework for distinguishing astrophysical environs and processes across the mass spectrum. 

The flow of trajectories toward the upper right of the plot shows an analogue of the stellar main sequence for black holes that we illustrated in Fig.~\ref{F:cartoon} and which we named the \emph{Cosmic Accretion} main sequence where black holes born in gaseous environments (such as high-redshift galaxies) can spend a significant amount of time. 
Utilizing realistic initial redshifts and better models for redshift evolution could lead to different clustering in the diagram. 
We note that canonical population III stars \cite{2023ARA&A..61...65K}, if included in Fig.~\ref{F:redshift}, would highly overlap the stellar-mass and intermediate-mass distributions with uncertain natal spins and evolve toward the supermassive regime 
through the \emph{Cosmic Accretion} main sequence; whereas supermassive population III stars \cite{2018MNRAS.474.2757H,2025MNRAS.536..851C} that form from gravitational collapse of atomically-cooled primordial halos would serve as a heavy seed source of supermassive black holes and possibly occupy the mass--spin gap in the lower right portion of Fig.~\ref{F:redshift}. Further populations can be added to the diagram to extend its utility for specific use cases, such as other direct collapse black hole formation possibilities or primordial black holes. 

\begin{figure*}
\centering
\includegraphics[width=\textwidth]{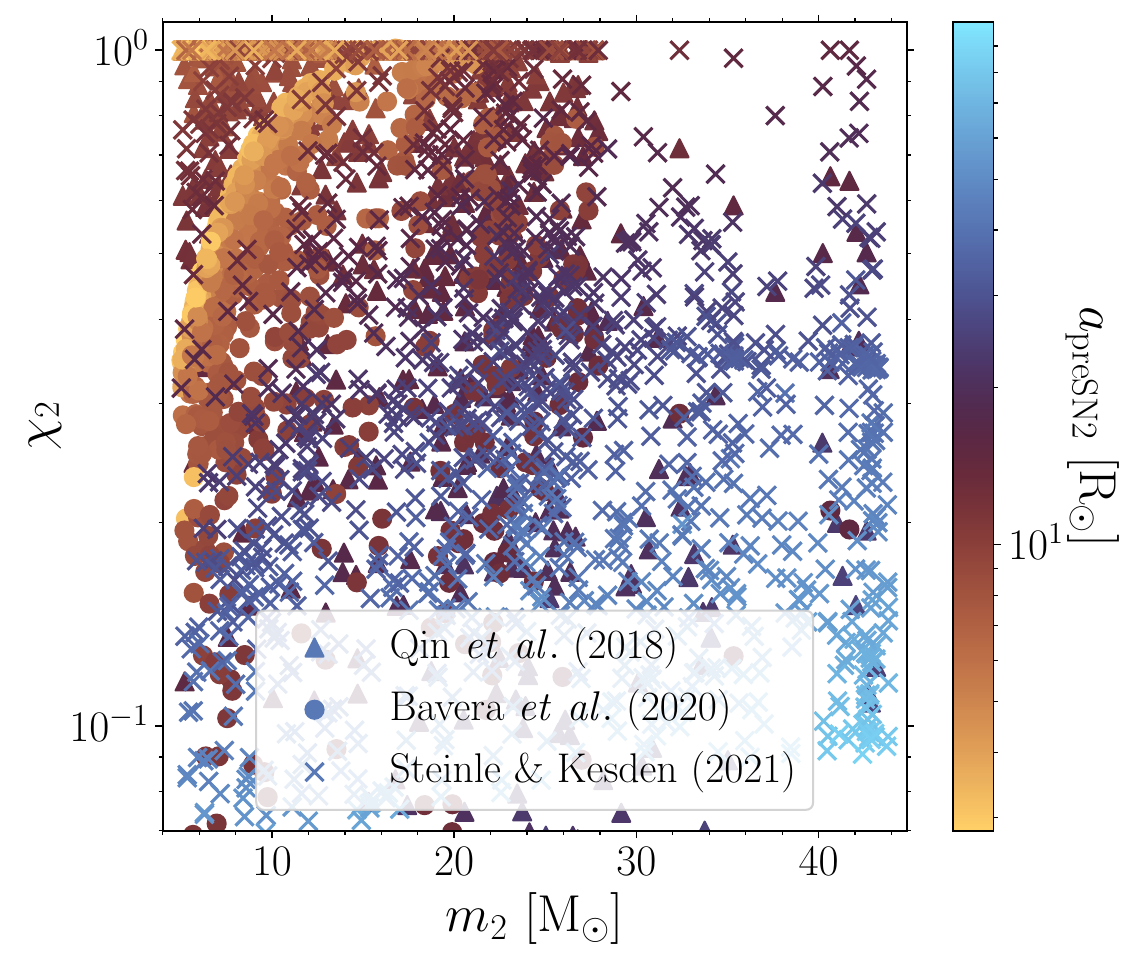}
\caption{
The predictions of three models for the spin magnitude of the Wolf-Rayet star tidally synchronized with a black hole companion in the isolated formation channel. 
The masses $m_2$ and spin magnitudes $\chi_2$ correspond to those of the secondary black holes (i.e., formed from the initially less massive star) in binaries from the COMPAS population \cite{COMPASdata} with color corresponding to the semi-major axis $a_{\rm preSN2}$ prior to its formation in a supernova explosion. 
The three spin models are indicated by triangles (Qin \textit{et al.}~(2018)), circles (Bavera \textit{et al.}~(2020)) and x's (Steinle \& Kesden~(2021)). 
The models of Qin \textit{et al.}~(2018) and Bavera \textit{et al.}~(2020) and of Steinle \& Kesden~(2021) are numerical and analytical 
adaptations of the model of \cite{1981A&A....99..126H}, respectively. 
} \label{F:COMPASmassspin}
\end{figure*}

Next, we examine a \texttt{COMPAS} stellar-mass black hole population, acquired from a public repository \cite{COMPASdata}\footnote{\href{https://zenodo.org/records/6346444}{COMPAS data on Zenodo}} which assumed all stars have the same metallicity $Z = 0.001$. 
Figure~\ref{F:COMPASmassspin} shows a subpopulation of black hole masses and spins where the Wolf--Rayet progenitor spins are evolved with three tidal synchronization models. 
These models all use the same underlying binary tidal evolution framework of \cite{1981A&A....99..126H} for Wolf–Rayet star tidal evolution in the case where the binary is composed of a black hole enforcing tides on its Wolf-Rayet companion in the isolated binary formation channel. But there are major differences in their implementations of that framework. 
Fig.~\ref{F:COMPASmassspin} shows the resulting dimensionless spin magnitudes $\chi_2$ of the black holes that formed from the initially less massive star as a function of their masses $m_2$. The points are color-coded by the orbital separation $a_{\rm preSN2}$ prior to the supernova explosion of the secondary. 
We compare three models\footnote{Two such models were already implemented into a \href{https://github.com/TeamCOMPAS/COMPAS/blob/dev/online-docs/pages/User\%20guide/Post-processing/notebooks/spin\_prescriptions/spin\_class.py}{COMPAS post-processing script}, which we have extended.} for obtaining the spin of a Wolf-Rayet star under tidal evolution: Qin \textit{et al.}~(2018) \cite{2018A&A...616A..28Q} in triangles, Bavera \textit{et al.}~(2020) \cite{2020A&A...635A..97B} in dots, and Steinle \& Kesden~(2021) \cite{2021PhRvD.103f3032S} in x's.  
The spin prescriptions are applied to the same underlying \texttt{COMPAS} binary population, i.e., the same masses and separations are used for each spin model, ensuring consistent initial conditions across models. All shown binaries are selected to be merging within a Hubble time, and have pre-SN2 separations less than $100\,R_\odot$ as the tidal synchronization timescale is longer for larger $a_{\rm preSN2}$. 
This highlights how differing assumptions about the final evolutionary phase of the progenitor binary are seen in the mass--spin diagram. 
The different assumptions result in different numbers of tidally synchronized Wolf-Rayet stars, and hence different numbers of points in the diagram, relative to the underlying total population: $\approx 50\%$, $\approx 60\%$, and $\approx 20\%$ for Qin \textit{et al.}~(2018), Bavera \textit{et al.}~(2020), and Steinle \& Kesden~(2021), respectively. For Bavera \textit{et al.}~(2020) and Steinle \& Kesden~(2021), we assume WR star spins are tidally synchronized with the orbital motion when the orbital period is less than 1 day (and in the case of Steinle \& Kesden~(2021) we find that using Eq. (13) of \cite{2021PhRvD.103f3032S} is a highly conservative limit for tidal synchronization), and for Qin \textit{et al.}~(2018) following their prescription. 

This comparison reveals interesting features of the mass--spin distribution expected from isolated binary evolution modeling. 
Binaries with tight orbits ($a_{\rm preSN2} \lesssim 10\,R_\odot$) tend to produce high-spin black holes ($\chi_2 \gtrsim 0.5$), while wider systems result in lower spins due to the assumption that inefficient angular momentum transfer during loss of the progenitor star's envelope will produce a low natal Wolf-Rayet spin. The different spin models produce different trends in the mass--spin plane, indicating that spin magnitude is sensitive to the adopted physics of tidal evolution. For example, the Steinle \& Kesden~(2021) model has smooth dependence on the binary separation modulated by dependence on the mass ratio, and produces more maximally spinning black holes whereas the other two models produce comparatively more slow rotators ($\chi < 0.1$). 

While a detailed comparison between these tidal models is beyond the scope of this work, we note that a key difference between the models is the use of analytic approximations in the model of Steinle \& Kesden~(2021) compared to the analytic fits in the models of Qin \textit{et al.}~(2018) and Bavera \textit{et al.}~(2020). The predictions from Steinle \& Kesden~(2021) fill the space more uniformly but are less accurate relative to the details of stellar evolution. 
For example, these models differ in the use of the (second order) tidal coefficient $E_2$ that appears in the expression for the synchronization timescale of the equilibrium tide whose complicated dependence on stellar structure and composition is difficult to capture. The models of Qin \textit{et al.}~(2018) and Bavera \textit{et al.}~(2020) (both based on precomputed grids from the MESA stellar evolution code \cite{2011ApJS..192....3P} assuming the tide only acts on the radiative envelope) use the same prescription for $E_2$ as given in Qin \textit{et al.}~(2018), while Steinle \& Kesden~(2021) use the prescription given by \cite{2008EAS....29...67Z} which is often implemented in rapid binary population synthesis codes such as StarTrack \cite{2008ApJS..174..223B} and COMPAS \cite{COMPAS2022}. 
Interestingly, these model predictions can share significant overlap, suggesting a testing ground for machine learning classification as we will see in Subsection~\ref{subsec:BHSpinModels}. 

Therefore, the mass--spin diagram has applicability to multiple important open problems in black hole astrophysics:  
\begin{itemize}
    \item In \textit{cosmological simulations} tracking supermassive assembly via galaxy mergers, the mass--spin history encodes growth mode and redshift evolution \cite{2025ApJ...985..220T}. 
    
    \item For \textit{intermediate-mass black holes}, observed at high redshift or in dwarf galaxies, the diagram can diagnose whether their mass--spin signatures follow stellar or supermassive-like growth. 
    
    \item \textit{Population synthesis studies}, such as those using \texttt{POSYDON} \cite{2024arXiv241102376A}, can populate the diagram with synthetic black hole data under different formation prescriptions and test classification boundaries by comparing with EM and GW observations. 
    
    \item \textit{Machine learning} trained on mass--spin data could identify formation possibilities and transitional or outlier objects that challenge existing classification schemes. 
\end{itemize}
While all are tantalizing prospects, here we focus on the last item to explore in the next section. 

We note that one can construct analogous diagrams, highlighting the ultimate flexibility of this framework, to adapt it to specific use cases. For example, one could transform the spin magnitude axis into a redshift axis given a cosmological model for the spin redshift-dependence for future gravitational-wave detectors. This diagram can also be extended to three dimensions by including a new axis or colormap for the spin orientation if the black holes exist in binaries, or relevant astrophysical parameters as in Fig. \ref{F:COMPASmassspin}. Other quantities that depend on the black hole mass(es) and spin(s), such as the recoil kick velocity in black hole formation from gravitational collapse or in black hole mergers of binary systems, would appear as contours on the mass-spin diagram. 
Even though the sensitivity of the LVK detectors vanishes beyond $z \approx 0.1$, the masses and spins and formation rates of stellar-mass black hole binaries depend on the redshift and metallicity dependent star formation rate evolution \cite{2023ApJ...957L..31F,2025CQGra..42e5009F}. 
The black hole spins are long expected to be key tracers of that evolution and to act as signposts for certain formation pathways and channels, and can thus be used as a proxy -or an intermediate- for the redshift in some cases such as strongly accreting populations, allowing for new comparisons between X-ray binaries and merging binaries \cite{2022ApJ...929L..26F,2022ApJ...938L..19G}. 
Future GW detectors will open the horizon to much higher distances, ie $z \lesssim 10$, indicating a need for new theoretical tools that encapsulate the wealth of scientific information these detectors will provide. The mass-spin diagram offers a natural framework to study the combined EM and GW black hole populations, as they generally originate in diverse environments and at varying redshifts, helping to empower future multi-messenger astrophysics studies of black holes of varying masses \cite{2024FrASS..1101792C}. Indeed, this multi-messenger concept can be extended to non-binary sources of GWs such as continuous wave emission from Galactic isolated neutron stars with LVK or ultra-compact binaries with LISA. 

\begin{figure*}
\centering
\includegraphics[width=\textwidth]{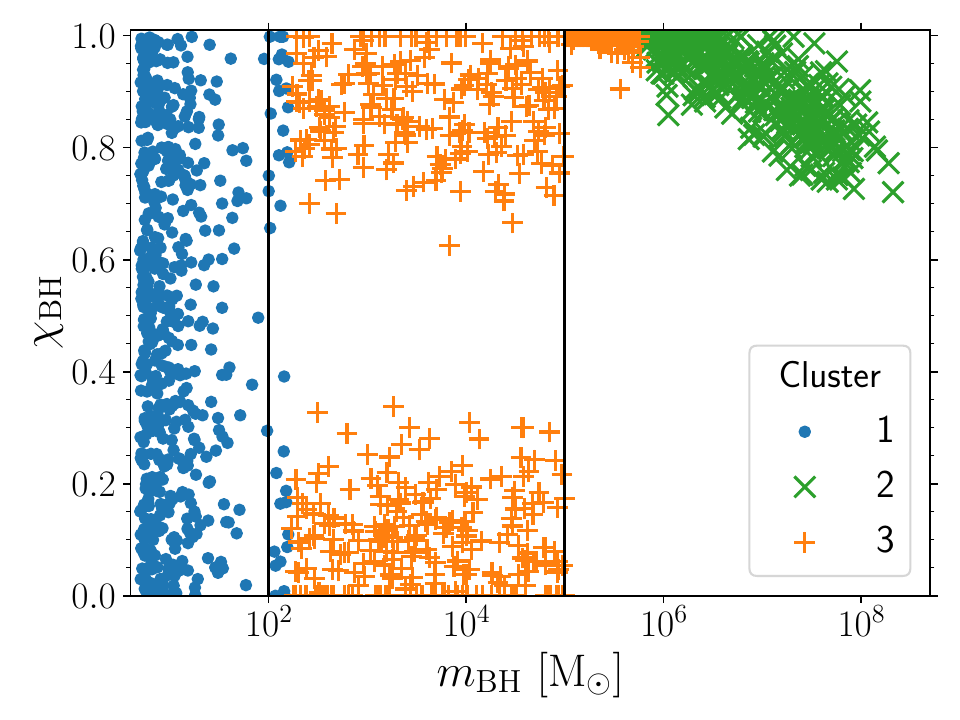}
\caption{
Unsupervised spectral clustering \cite{2011JMLR...12.2825P} on the natal black hole mass--spin correlations from Fig.~\ref{F:redshift}. The resultant clusters are shown by blue circles, red pluses, and green x's. 
Black vertical lines show the two mass boundaries imposed in the true dataset between the stellar-mass, intermediate-mass, and supermassive regimes. 
} \label{F:continuum}
\end{figure*}

\section{Classifying black holes}
\label{sec:Classifying}

Machine learning algorithms require comprehensive datasets for training otherwise the classification predictions degrade. This can happen in at least two main ways which limit their ability to generalize predictions: either the dataset sparsely fills the total parameter space volume of the system, or the dataset exhibits degeneracies such as overlapping contributions from multiple components. In this section, we will explore both cases (i.e., the former in Subsec.~\ref{subsec:MassContinuum}, and the latter in Subsec.~\ref{subsec:BHSpinModels}) in the context of black-hole binary populations. 
We construct these datasets from toy-model data or from the raw output of realistic simulations in order to test this machine learning framework in a controlled setting, and we leave application to real-world datasets (i.e. from gravitational and electromagnetic observations) to a future study. 

\subsection{The Black Hole Continuum}
\label{subsec:MassContinuum}

Intermediate‑mass black holes (IMBHs; $\approx 10^2\,\Msol$ to $10^5\,\Msol$) occupy a contentious region between well-established stellar‑mass and supermassive regimes. EM candidates in ultra‑luminous and hyper-luminous X‑ray sources, dynamical  globular clusters, or dwarf galaxies are numerous but have yet to be confirmed confidently \citep{2020ARA&A..58..257G}. Meanwhile, recent GW events with LVK (e.g., GW190521, GW200220, GW231123) concretely confirm the presence of lower-end IMBHs, offering glimpses into this apparent mass gap.

The traditional tripartite division—stellar, intermediate, supermassive—is a human‑constructed taxonomy based on discovery methods and observational biases: EM surveys select high‑luminosity or high‑mass remnants, and GW detections of stellar-mass and intermediate mass black holes sample different redshift and mass ranges. For example, pulsar timing arrays are more sensitive to high-mass, low redshift supermassive black-hole binaries, and LISA will be sensitive to lower mass black holes to high redshift. These selection effects correlate with mass--spin trends inherently tied to astrophysical origins. In this subsection we examine whether machine learning with \textsc{Scikit-learn} \cite{2011JMLR...12.2825P} can discover these human‑assigned distinctions using only mass and spin features. 

First, we generate a toy dataset comprising two populations that reflect observed masses and spins (see Table~\ref{Table:summary_BHs}): stellar‑mass black holes with spin magnitudes uniformly distributed between 0 and 1, and supermassive black holes with spins uniformly between 0.9 and 1. Figure \ref{F:IMBHgap} shows how a Gaussian mixture model (left panel) and k-means (right panel) for clustering on this bimodal distribution successfully separates stellar and supermassive populations due to the simple mass-spin correlations. We perform this clustering in log-mass space to ensure proper feature scaling. This hints toward a potential for machine learning methods to reproduce—and ultimately refine—existing classifications, anticipate new categories as datasets grow, and identify biases inherent in human-based taxonomy. 

Building on this preliminary test, next we construct a more astrophysically realistic black hole continuum in Figure \ref{F:continuum}. Here, we sample three initial mass functions (the same ones used to population Fig.~\ref{F:redshift}) corresponding to stellar, intermediate-mass (uniform), and supermassive populations, and we assign each regime spin magnitude distributions of increasing concentration: uniform in [0,1] for stellar, bimodal for intermediate, and high spins ($\chi\gtrsim 0.8$) for supermassive black holes. This theoretical dataset can challenge unsupervised learning algorithms to discover three clusters without knowledge of their labels.   

Spectral clustering (with nearest neighbor affinity), which operates on the full mass--spin manifold, can struggle to accurately recover the tripartite division, as shown by the three color-marker pairs in Fig.~\ref{F:continuum} where the vertical black lines indicate the true mass boundaries assumed in constructing the theoretical dataset. The misclassification of edge cases reflects the spin distinctions we imposed, but more importantly it demonstrates that such algorithms can detect latent structure in complex data which may arise from selection effects or data analysis assumptions. 
Also, it provides further motivation for the development of a black hole continuum mass function, which is an active area of research \cite{2022ApJ...924...56S} and would present an interesting challenge for these strategies. 
While the IMBHs are composed of two subclasses with respect to the spin axis, the algorithm did not identify them as separate clusters, likely due to our assumption of three classes in the learning. In future work, we will explore how this depends on the number of classes one assumes in the unsupervised learning and its implications for future observations of IMBHs. 
We note that this classification was performed without latent space embedding prior to clustering, implying that an analysis which includes latent space transformation may perform even better, which we leave to future work.

\subsection{Deep Learning of Stellar-mass Black Holes}
\label{subsec:BHSpinModels}

The analysis in the previous section confirms that machine learning tools can recognize broad mass--spin categories when they are not highly overlapping in the mass--spin parameter space. But many potential applications of the mass--spin diagram will involve data with overlapping black hole subpopulations that do not occupy distinct regions of the parameter space and instead exhibit distinct mass-spin correlations over similar ranges of mass and/or spin. This case will be particularly challenging for unsupervised classification due to the degenerate latent space and, as we shall see below, provides strong motivation for development of advanced machine learning techniques for embedding astrophysical latent information and analyzing it with the mass--spin diagram. 

We explore classification of three such datasets for stellar-mass binary black hole systems where each is composed of two classes: 
\begin{enumerate}
    \item black hole spin prescriptions from an isolated binary population model; 
    \item black hole mass--spin correlations from isolated binary and dynamical cluster population models; 
    \item physically motivated $\chi_{\rm eff}$--$q$ correlations for the isolated and dynamical cluster formation channels. 
\end{enumerate} 

In case 1., we take the data plotted in Fig.~\ref{F:COMPASmassspin} by taking the secondary mass $m_2$ and spin magnitudes $\chi_2$ from the Bavera \textit{et al.}~2020 and Steinle \& Kesden~(2021) models as these produce the least overlapping regions in Fig.~\ref{F:COMPASmassspin}. We limit spin magnitudes to between $10^{-4}$ and $1.0$ to remove non-rotating systems, apply a log transform, and standardize the masses and spins separately via a Gaussian offset $\mathcal{N}(\log_{10}~x, 0.01)$ where $x$ is the mass or spin, and pairing each black-hole binary with corresponding true model labels. 
The standardized mass and spin features are combined into two-dimensional feature vectors $(\mathrm{m}, \mathrm{\chi})$ for each sample with corresponding integer true class labels for each model, producing the final feature--label (i.e., input--output) datasets. 

Similar steps are followed for cases 2. and 3. 
In case 2, the main difference is using two different mass--spin datasets: one from \texttt{COMPAS} \cite{COMPASdata} with the Steinle \& Kesden~(2021) spin prediction and one produced with the \texttt{Rapster} code \cite{2024PhRvD.110d3023K} with default settings and $N = \num{100e+6}$ initial stars, resulting in two  black-hole mass and spin distributions that overlap highly for mass $m_2 \lesssim 43\ \Msol$, above which only dynamically formed black-hole binaries form with spins concentrated in the range $\chi_2 \sim 0.5$ to $0.7$, consistent with GR remnant-spin fits. 

Instead of the output of simulations, for case 3 we construct distributions of mass and spin to simulate two theoretical correlations in the $\chi_{\rm eff}$--$q$ parameter space corresponding to isolated binaries or dynamically formed binaries. First we sample the mass ratio $q$ uniformly in $(0.2, 0.9998)$ and total mass $M$ uniformly in $(30, 100)\,M_{\odot}$, from which the component masses $m_1$ and $m_2$ are derived. 
For the dynamical-cluster binaries, spin magnitudes $\chi_1$ and $\chi_2$ are set to be relatively high ($\sim 0.7$) and decreasing linearly with $q$, and spin-orbit misalignment cosines $\cos\theta$ are drawn uniformly from $[-1, 1]$. The effective spin $\chi_{\mathrm{eff}}$ is computed from these quantities.  
For the isolated binaries, $\chi_{\mathrm{eff}}$ is computed from spin magnitudes that are lower on average and given by $\mathcal{N}(0.5, 0.1)$, and spin-orbit misalignments $\cos\theta$ that are mostly aligned and modeled as decreasing linearly with $q$ and with spread given by $\mathcal{N}(0.5, 0.1)$. 
Another main difference for case 3 is that the latent space is defined by $\chi_{\rm eff}$--$q$ instead of the individual masses and spins, incorporating more information than using the binary masses and spins. 

Using \texttt{PyTorch} \cite{paszke2017automatic} we implement and benchmark a neural network architecture which imposes a regularized latent structure, and we perform clustering with \texttt{scikit-learn} \cite{2011JMLR...12.2825P}. 
We do this in a modular, end-to-end machine learning framework to extract latent structure in simulated black hole populations and perform classification \footnote{Codes used in this study will be made \href{link}{publicly available} upon publication}. The workflow follows a standard pipeline, where raw simulation outputs are extracted, processed, embedded into a compressed latent space with \textsc{PyTorch} neural net architecture, and clustered with the \textsc{scikit-learn} machine learning suite. We use an unsupervised classification scheme via a variational autoencoder (VAE) and k-means clustering, extending it to a semi-supervised classification framework, and a random forest supervised classifier. Raw data are split into 60\% training, 20\% test, and 20\% validation sets containing black hole masses, spin magnitudes, and true class labels. As the performance is the same to within 10\% between the training, test, and validation datasets, we only show the training results for brevity. 
All three datasets contain 100,000 black holes in each class, but results in Table~\ref{Table:summary_stats_case1} are similar for 10,000 black holes each. 

The VAEs treat the encoder outputs as parameters of a Gaussian distribution to build a representation of the embedding space. 
In training the VAEs (i.e., one for each of training, test, and validation datasets), we deploy the full machine learning stack using \textsc{PyTorch}'s neural network autograd engine, i.e. managing backpropagation across both encoder and decoder components, and automatic differentiation training with the Adam optimizer (fiducial settings: 100 epochs, learning rate 0.001, batch size of 512) and early stopping criteria which combines reconstruction loss with a Kullback–Leibler divergence penalty. 
Unsupervised clustering is performed with the k-means algorithm on the means of the latent variables. 
Since for our three theoretical datasets the true classes are known, we also employ a supervised random forest classifier via \texttt{Sci-kit Learn} on the VAE latent space and perform clustering with 200 estimators. 
With both classifiers, output prediction of the classes is aligned (i.e., which cluster corresponds to which spin or formation model) to the true underlying distribution after clustering via linear sum assignment in \textsc{SciPy} \cite{2020SciPy-NMeth} to ensure self-consistent labeling for comparisons.   

\input{table2.tex}

For both the unsupervised and supervised classification methods, the clustering results are interpreted through comparison with ground-truth labels to compute the accuracy; the F1-score which measures a model's ability to correctly identify positive cases and its overall accuracy; and the adjusted Rand index (ARI) and normalized mutual information (NMI) which measure how different the learned latent space correlations are from a random assignment of labels to measure the learning of the latent space (0 means completely random and 1 means the latent space recovers the true correlations). 

The classification and benchmarking results for the three datasets are shown in Table~\ref{Table:summary_stats_case1}, where we only report the training results for brevity as the test and validation results are similar (i.e. $\lesssim 10$\% difference). 
The unsupervised k-means clustering is unable to correctly learn the highly overlapping mass--spin distributions. 
This is demonstrated by the $\approx 0$ ARI and NMI indicating ineffective latent space learning. 
The consistent F1-score of $\sim 0.5$ for k-means across different hyperparameter settings combined with the near-0 ARI and NMI indicates that the k-means approach ends up with a random $\sim 50/50$ clustering assignment. 
We also confirm independently that the ARI and NMI are 0 ($\sim 0.3$) when unsupervised (supervised) clustering is performed directly on the transformed raw input data rather than on the latent representation space. 
We also performed this analysis with a Gaussian mixture model instead of k-means and it performed similarly. 

\begin{figure*}
\centering
\includegraphics[width=\textwidth]{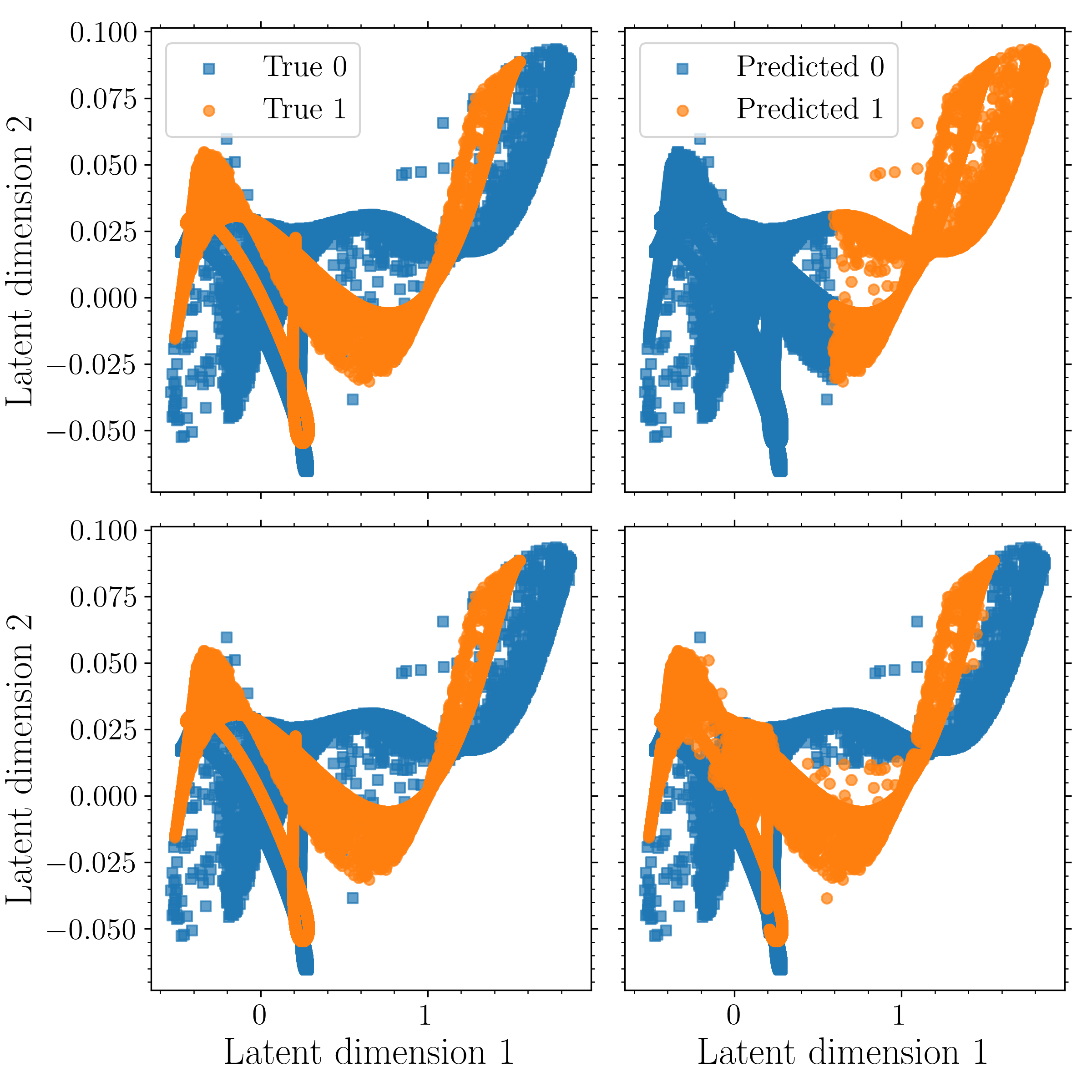}
\caption{
Latent space representation of dataset 1. in Table~\ref{Table:summary_stats_case1} with a variational autoencoder from PyTorch \cite{paszke2017automatic}. The number of latent dimensions equals the number of raw input dimensions (black hole mass and spin). The true latent space is plotted for clarity in both panels of the left column, and the predicted latent space is plotted in the panels of the right column for the unsupervised k-means (top) and the supervised random forest (bottom).  
} \label{F:latent}
\end{figure*}

Figure~\ref{F:latent} demonstrates the difference in clustering in the trained VAE latent space between the unsupervised (top row panels) and supervised (bottom row panels) clustering of dataset 1., where the true and predicted data correspond to the left and right columns, respectively. Blue circles and orange squares correspond to the two spin models of Bavera \textit{et al.}~(2020) and Steinle \& Kesden~(2021), respectively. This visually confirms the challenge of latent space learning for the unsupervised approach. 
The success of the supervised random forest classifier (i.e., $\sim90\%$ accuracy) is shown in the bottom-row panels of Fig.~\ref{F:latent}, as the two predicted clusters are closely reproduced in the right panel with respect to the true clusters in the left panel. The dataset 1. has more distinct classes compared to datasets 2. and 3., and the spins of the two classes in dataset 1. correspond to the same black hole masses but in datasets 2. and 3. the spins of the two classes have different corresponding masses; both of these features contribute to the random forest performing better on dataset 1.  

While the unsupervised approach will require further investigation in future work, the success of the supervised random forest classifier indicates that a semi-supervised approach might show intermediate performance. 
Thus, we expanded the k-means pipeline for semi-supervised classification in which we assume some fraction has known labels and the training for the unlabeled fraction is the same as in the unsupervised pipeline but training for the labeled fraction employs an additional penalty in the loss function where points farther away from each other in the parameter space have less weight via cross-entropy loss in \textsc{PyTorch}. 
The result of the semi-supervised approach is sensitive to the initialization of random states in several components of the pipeline. 
We ran the semi-supervised k-means pipeline with 5 different initializations and a labeled fraction of 0.5. 
The resultant best semi-supervised clustering had accuracy $\approx$ 0.6, f1-score $\approx$ 0.58, and NMI and ARI $\approx$ 0.45. 
Comparing with the fully unsupervised case, this demonstrates that the low-dimensional latent space transformation can improve clusterability, but a full analysis is beyond the scope of the present study. 

The use of uncertainty in the event parameters is a key area of future development, suggesting the use of methods like Bayesian Neural Networks that can quantify uncertainty in training and prediction. 
Such an analysis can be paired with hierarchical inference output \cite{2021ApJ...922L...5C,2022PhRvD.106j3013M}, allowing the ML algorithms to operate directly on the posteriors rather than summary statistics (i.e. medians), enabling better utilization of astrophysical information for latent space learning. 
Our results indicate that combining the mass--spin diagram with machine learning contains potential for many applications in studying astrophysical black hole populations where distinguishing formation channels which is a great open problem in gravitational-wave astronomy, see 
as examples \cite{2016ApJ...830L..18B,2021ApJ...910..152Z,2013PhRvD..87j4028G,2016ApJ...832L...2R,2017MNRAS.471.2801S,2023ApJ...955..127C,2024PhRvL.133e1401L}. 
This motivates development of new unsupervised and semi-supervised learning methods, and fine-tuning of the approaches attempted here, for challenging astrophysical datasets from future GW and EM detectors, which we leave to a followup study \cite{Steinle2025}. 
Another avenue of future work will use more realistic combined populations of black-hole binary formation channels in a cosmological setting, i.e., the redshift evolution across the mass--spin diagram.

\section{Summary}
\label{sec:Sum}

We present a new framework for visualizing and interpreting black hole populations with the mass--spin  diagram, analogous to the classic HR diagram in stellar astrophysics. While the HR diagram has long served as a cornerstone in our understanding of stellar evolution, a similarly intuitive and physically motivated diagnostic tool for black holes can be found in the mass--spin diagram. Capturing imprints of mass and angular momentum evolution, one can examine the time or redshift evolution and compare formation possibilities. 

First we examined the mass--spin diagram in the context of the three black hole mass regimes: stellar-mass black holes formed from gravitational stellar core collapse, intermediate-mass black holes possibly formed through runaway collisions or direct collapse, and supermassive black holes from light and heavy seeds. 
We approximate redshift evolution using physically motivated mass and spin prescriptions, incorporating both super-Eddington accretion and stochastic processes. These black hole trajectories serve as tracks in the mass--spin space, shaped by their accretion history and seed properties. For instance, stellar-mass BHs can spin up efficiently via fallback or disk accretion, while intermediate-mass black holes undergo diverse accretion and merger episodes, and supermassive black holes experience coherent spin-up in the early Universe and hierarchical galaxy mergers. 
The mass--spin trajectories carry a fossil record of evolutionary mechanisms, analogous to stellar tracks and clusters in the HR diagram. 

Additionally, we provide a zoomed-in view on the stellar-mass black hole regime in a mass--spin diagram constructed with data from the COMPAS binary evolution code. We post-process the black hole data to obtain their spins with three models for the evolution of Wolf-Rayet tidal evolution in progenitor binaries of black-hole mergers. 
The mass--spin diagram exposes differences in modeling assumptions motivating development of comprehensive and consistent spin evolution models. This will be most useful in the era of future terrestrial GW detectors, where catalogs of black hole binaries are expected to expand by at least 100 times in size \cite{CE2021,ET2025}, consistent with the size of the populations utilized here. 

We apply unsupervised and supervised machine learning for classification of a series of realistic and progressively challenging datasets. 
First we show how simple unsupervised algorithms can recover features in the mass--spin space, such as the intermediate-mass gap with k-means clustering and commonly used mass boundaries in a mass continuum with spectral clustering. 
Moving beyond simple clustering, we employ deep learning with a variational autoencoder (VAE), a type of neural network, to learn low-dimensional latent embeddings of three synthetic datasets for black hole masses and spins. After training the VAEs on standardized features using \texttt{PyTorch}, we extract the 2D latent spaces and apply an unsupervised algorithm, k-means clustering, and a supervised algorithm, random forest, for machine learning classification. 
In all three cases, the k-means approach does not learn the latent space representation and fails to accurately cluster the black holes according to their mass--spin correlations. This is clearly shown in Fig.~\ref{F:latent} where the unsupervised latent space learning is shown in the top row panels compared to the supervised learning in the bottom panels for true (left column panels) vs predicted (right column panels). 
The random forest performs best for the first dataset, which is composed of two classes of black hole spins from models with differing assumptions about Wolf-Rayet tidal evolution, with near perfect accuracy and high ARI/NMI. 
This is because the two classes exhibit more distinct correlations in the mass--spin space compared to the other two datasets which have highly overlapping classes. In the case of dataset 2, this is simply the outcome of comparing raw simulation output from two state-of-the art models of black-hole binary formation channels; in the case of dataset 3, this is by construction to approximate an observed correlation in LVK data.   

In this work, the number of clusters in the unsupervised classification of Fig.~\ref{F:continuum} was fixed to three to reflect the canonical stellar-mass, intermediate-mass, and supermassive regimes and to test whether the algorithm could recover these established divisions. More generally, however, the optimal number of clusters is not known \emph{a priori} in unsupervised learning and can be estimated through data-driven criteria such as the elbow method, silhouette score, or inspection of the Laplacian eigenspectrum gap in spectral clustering. Future applications of this framework to observational datasets will employ such methods to algorithmically infer the population structure. Likewise, the dimensionality of the variational autoencoder latent space, as shown in Fig.~\ref{F:latent} and which was set to two for direct comparison with the mass--spin diagram, is related to the dimensionality of the mass--spin diagram and can be treated as a tunable hyperparameter. Exploring higher-dimensional latent representations, including additional astrophysical parameters such as spin orientation, redshift, or accretion rate, will form part of future work aimed at improving cluster separability and physical interpretability.

These results show the utility of the mass--spin diagram for quantifying differences between spin evolution models and distinguishing formation channels of black-hole binaries. Our application of machine learning classification to the diagram reveals interesting challenges, motivating the development of advanced techniques to learn astrophysical datasets. Future work will identify how the motion of black holes across the mass--spin diagram can aid in their classification.

\section{Discussion}
\label{sec:Disc}

Out results have multiple implications. First the mass--spin diagram provides a visual and statistical framework for classifying BHs across all mass scales, allowing direct comparison of theory and data in a physically meaningful space. 
Machine learning of gravitational-wave populations in the diagram is useful for studying formation possibilities. 
This framework can be improved by incorporating merger products, eccentricity, environmental gas effects, and host galaxy priors into the mass--spin space. As observed catalogs grow in size and fidelity, these methods will become increasingly useful for inferring formation histories of black holes.
In the HR diagram, stars on the main sequence occupy the primary band defined by core hydrogen burning in hydrostatic equilibrium, a fundamental and long‐lived phase. 
We suggest that analogues for black holes may be tied to mass--spin growth tracks, such as prolonged, coherent accretion episodes, as the black hole evolves in redshift.  
Numerous theoretical and cosmological simulation studies support the universal nature of accretion-driven mass and spin growth \cite{2019ApJ...883...76R,2015SciA....1E0686S,2025arXiv250619166A,Yang22}. These simulations imply that gas accretion dominates supermassive black hole growth in the high-redshift universe allowing for efficient spin-up, while galaxy mergers dominate at lower redshifts, possibly providing random spin orientations relative to the host galaxy; together, these create mass--spin tracks for intermediate and supermassive black holes.  
A recent study found a three-phase evolution, depending on the mass and redshift of the black hole, with predictions for their spin magnitudes and orientations relative to their host galaxies \cite{2024A&A...686A.233P}. 
General relativistic magneto-hydrodynamic simulations show the interplay of accretion physics and generic spin evolution in gravitational-wave events \cite{2024PhRvD.109d3004C}. 

Analogues to the stellar main sequence, which is the band in the HR diagram where stars spend the most time, can be found in the mass--spin diagram. 
An example we explore is \emph{Cosmic Accretion}, a main sequence 
due to prolonged episodes of accretion 
across diverse formation environments and pathways, seen as mass--spin snapshots of all black holes within a certain redshift bin. 
Determining these snapshots accurately would require realistic models of the origin and evolution of black hole populations across mass and redshift ---for stellar-mass black holes this is tied to metallicity-dependent star formation history and for more massive black holes is tied to the cosmological evolution of galaxy clusters. 
Mechanisms that drive mass and spin evolution possibilities of black holes create tracks through the mass--spin diagram which can be marked by continuous or discontinuous transitions, such as from accretion or mergers, respectively.  
Deviations from these tracks, such as mechanisms like low-spin natal black holes, hierarchical mergers, or chaotic accretors, can play the role of post-main-sequence stars or blue-loop outliers. 
Black hole binaries in isolation will be static on the diagram unless they merge, resulting in a merger-tree structure analogue type of main sequence, where black holes appear and disappear as merger products or remnants and tend to pile-up in regions of the diagram according to merger-dependent mass and spin distributions. Ultimately, the many main sequences would be connected through cosmological evolution of the Universe. 

Our study yields three main new testable predictions:
\begin{itemize}
    \item the existence of mass--spin gaps in the diagram resulting from mass--spin correlations. 
    \item the existence of black hole main sequences tied to formation channels; channels with accretion will have distinct motions on the mass--spin diagram compared to channels with mergers. 
    \item the application of machine learning to classify black hole populations within and across mass regimes according to observed and theoretical spin prescriptions and correlations. 
\end{itemize}
For example, in the stellar-mass regime, it is known that EM and GW observed populations differ in metallicity and redshift \cite{2022ApJ...929L..26F,2022ApJ...938L..19G}, suggesting that machine learning classification can aid in categorizing and generalizing information of these distinctions from future datasets. 
There are important caveats to our analysis. We use various theoretical datasets to explore this new application of machine learning, but the application to real data will encounter new challenges not revealed in the present study. 
These datasets range from toy models of theoretical expectations and of observed correlations in real data to raw simulation output of realistic models. These products of theoretical formation channels of black holes play a crucial role in constructing useful training datasets for the machine learning algorithms. 
Also, we did not explore the motion of black holes across the diagram in our machine learning analysis - utilizing such black hole tracks will open a new area of application of the techniques demonstrated here. 

The future of GW astronomy is bright with the construction and planning of many detectors across the GW frequency spectrum, with the Square Kilometer array  \cite{2009pra..confE..58L}, MeerKAT \cite{2025MNRAS.536.1467M}, and DAS-2000 \cite{2019BAAS...51g.255H} in the nano-Hz regime, astrometric facilities such as Theia \cite{2018FrASS...5...11V} in the micro-Hz regime, LISA \cite{2017arXiv170200786A}, TianQin \cite{2016CQGra..33c5010L}, and Taiji \cite{2018arXiv180709495R} in the milli-Hz regime, and Cosmic Explorers \cite{CE2021} and the Einstein Telescope \cite{ET2025} in the Hz to kilo-Hz regime. Interestingly, one can perform similar analysis as done here on a diagram defined by GW-frequency and black hole spin to compare the spins of binary mergers as a function of GW frequency. 
Multi-band studies can compare the populations detectable by pulsar timing arrays and LISA (supermassive black holes) with those detectable by LVK, Cosmic Explorer and Einstein Telescope (stellar-mass black holes), by their spin evolution and with both sets of detectors peering into the intermediate-mass black hole regime. 
Ultimately, the potential for multi-messenger applications of the mass--spin diagram are also plentiful with the development of future X-ray facilities such as AXIS \cite{2023SPIE12678E..1ER} which will provide unprecedented measurements of accreting systems. The classification of joint EM and GW populations, as done in Fig.~\ref{F:continuum}, will be an important direction of future work in multi-messenger astronomy.

\acknowledgements
The Authors thank Nicole Ndegwa for support, inspiration, and perspicacious suggestions for visualizations. 
We also thank Davide Gerosa for helpful feedback on machine learning methods, Michael Kesden for inspiring early comments, 
Austin MacMaster for insightful discussions, and the anonymous referee for constructive recommendations. 
N.S. and S.S.H. are supported by the Natural Sciences and Engineering Research Council of Canada through the Canada Research Chairs and Discovery Grants programs. 
This research was enabled by use of the GREX cluster (https://um-grex.github.io/grex-docs/friends/alliancecan/) and the Digital Research Alliance of Canada (alliancecan.ca). 
Simulations in this paper made use of \textsc{PyTorch} \cite{paszke2017automatic}, \textsc{Scikit-learn} \cite{2011JMLR...12.2825P}, \textsc{SciPy} \cite{2020SciPy-NMeth}, \textsc{NumPy} \cite{numpy}, \textsc{Pandas} \cite{pandas1,pandas2}, \textsc{h5py} \cite{h5py}, \textsc{matplotlib} \cite{matplotlib}, and the COMPAS rapid binary population synthesis code (version 02.42.01) which is freely available at \href{http://github.com/TeamCOMPAS/COMPAS}{http://github.com/TeamCOMPAS/COMPAS}.

\bibliography{BHclass}

\end{document}

%% file: journal_macros.tex
\def\araa{{Ann.~Rev.~Astron.~Astrophys.}}
\def\apj{{Astrophys.~J.}}\def\apjl{{Astrophys.~J.~Lett.}}
\def\apjs{{Astrophys.~J.~Supp.}}
\def\ao{{Appl.~Opt.}}
\def\apss{{Astrophys. Space Sci.}}
\def\aap{{Astron.~Astrophys.}}

\def\mnras{{Mon.~Not.~R.~Astron.~Soc.}}             

\def\nar{{New A. Rev.}}

\def\prd{{Phys.~Rev.~D}}

\def\prl{{Phys.~Rev.~Lett.}}

\def\ssr{{Space~Sci.~Rev.}}

\def\physrep{{Phys.~Rep.}}

%% file: table.tex
\newcolumntype{C}[1]{>{\centering\arraybackslash}p{#1}}

\begin{table*}[htbp]
\caption{Summary of the three main mass regimes/classes of black holes from current electromagnetic (EM) \cite{2020mbhe.confE..28B,2013CQGra..30x4004R} and gravitational-wave (GW) \cite{2025arXiv250818082T} observations. Only stellar-mass black holes have been observed via both EM and GW messengers. The corresponding observed ranges of mass and spin measurements of each regime are indicated along with the methods for obtaining them: 
X-ray reflection spectroscopy (or iron-line method), spectral continuum fitting, reverberation mapping, and stellar dynamics.}
\label{Table:summary_BHs}
\begin{tabular}{C{1.25in}|C{0.7in}C{.9in}C{0.9in}C{3in}}
 \hline
 BH class/name & Messenger & Mass range $\Msol$ & Spin range & Observation methods  \\
\hline
\hline
\multirow{2}{*}{Stellar-mass} & EM & $\approx 5 - 20$ &  $\approx 0.1-0.98$  & reflection, continuum, dynamics  \\
 & GW & $\approx 5 - 250$ & $\approx 0 - 0.998$ & unmodeled and modeled searches  \\
\hline
 & & & \\
Intermediate-mass  & EM & $\approx 10^2 - 10^5$ &  ? & reflection, continuum, reverberation, dynamics   \\
\hline
& & & \\
Supermassive & EM & $\gtrsim 10^6$ &  $\approx 0.6 - 0.99$ & reflection, continuum, reverberation, dynamics     \\ \hline
\end{tabular}
\end{table*}

%% file: table2.tex
\newcolumntype{C}[1]{>{\centering\arraybackslash}p{#1}}

\begin{table*}[htbp]
\caption{Performance metrics of clustering with an unsupervised classifier (k-means) and a supervised classifier (random forest) on the latent space representation from a variational autoencoder (VAE) for distinguishing the mass--spin correlations of the three black hole binary datasets described in the text. The metrics utilized here are the accuracy (i.e. the fraction of correctly labeled black holes), the adjusted Rand index (ARI), normalized mutual information (NMI), and F1-score.}
\label{Table:summary_stats_case1}
\begin{tabular}{C{1.5in}|C{1in}|C{1.25in}|C{0.7in}|C{0.7in}|C{0.7in}|C{0.7in}}
\toprule
Dataset & Type & Classifier & Accuracy & ARI & NMI & F1-score \\
\hline
\multirow{2}{*}{1. Isolated binary tides} & unsupervised &  KMeans & 0.51  & 0.01  & 0.01 & 0.50 \\
 & supervised & Random Forest & 0.95 & 0.9 & 0.8 & 0.95 \\
\hline
\multirow{2}{*}{2. Isolated vs dynamical} & unsupervised &  KMeans & 0.6 & 0.01  & 0.01 & 0.6 \\
& supervised & Random Forest & 0.8 & 0.3  & 0.3 & 0.8  \\
\hline
\multirow{2}{*}{3. $\chi_{\rm eff}$--$q$ correlation} & unsupervised &  KMeans & 0.56  & 0.01  & 0.01 & 0.55 \\
 & supervised & Random Forest & 0.8 & 0.46 & 0.4 & 0.82 \\
\hline
\end{tabular}
\end{table*}